\def\ifmath#1{\relax\ifmmode #1\else $#1$\fi}%
\def\rd{\ifmath{{\mathrm{d}}}}
\def\rK{\ifmath{{\mathrm{K}}}}
\def\rp{\ifmath{{\mathrm{p}}}}
\def\rq{\ifmath{{\mathrm{q}}}}

\def\rs{\ifmath{{\mathrm{s}}}}
\def\rS{\ifmath{{\mathrm{S}}}}

\def\ru{\ifmath{{\mathrm{u}}}}
\def\rZ{\ifmath{{\mathrm{Z}}}}
\def\jet{\ifmath{{\mathrm{jet}}}}
\def\cut{\ifmath{{\mathrm{cut}}}}

\documentclass{ws-p8-50x6-00}

\begin{document}

\title{Recent Results on Particle Production from OPAL}

\author{J. H. Vossebeld}

\address{CERN, CH - 1211 Geneva 23, Switzerland\\E-mail: Joost.Vossebeld@cern.ch}


\maketitle

\vspace{-2.4in}
\begin{flushright} 
OPAL CR-459\\
hep-ex/0101009
\end{flushright}
\vspace{2.05in}

\abstracts{  
Three recent OPAL studies are presented in which the fragmentation 
process in quark and gluon jets and in identified up, down and strange 
flavour jets is studied. 
The first is a measurement of charged particle, 
$\pi^0$, $\eta$ and $\rK^0$ multiplicities in quark and gluon jets. 
No evidence is found for a particle-species dependent multiplicity 
enhancement in gluon jets. 
In another study, leading $\pi^\pm$, $\rK^\pm$, $\rK^0_\rS$, $\rp(\bar{\rp})$ 
and $\Lambda(\bar{\Lambda})$ rates have been measured in up, down 
and strange flavour jets. The results confirm the leading particle effect 
in the fragmentation of light flavour jets. In addition, a direct 
determination of the strangeness suppression factor has been performed, 
yielding $\gamma_\rs~=~0.422~\pm~0.049~({\rm stat.})~\pm~0.059~({\rm syst.}).$
In a third study, mean charged particle multiplicities were measured for  
up, down and strange flavoured $\rZ^0$ decays and found to be identical 
within the uncertainties of the measurement, as expected from the flavour 
independence of the strong interaction.
}

\section{Introduction}
The large number of hadronic $\rZ^0$ decays collected by the LEP experiments at
$\sqrt{s}=M_{\rZ^0}$ allow for detailed studies of the dynamics of the strong 
interaction. In three recent OPAL studies presented in this paper, 
production rates of various hadron species in the fragmentation of quark 
and gluon jets and of identified up, down or strange flavour jets have been 
studied. The results constitute a detailed test of models describing the 
fragmentation of partons in to hadrons in the final state.

\section{\boldmath $\pi^0$, $\eta$, $\rK^0$ and charged multiplicities in quark and gluon jets}

Because of the colour enhancement of the gluon-gluon coupling with 
respect to the quark-gluon coupling, the particle multiplicity in gluon 
jets is higher than in quark jets. As this multiplicity enhancement 
appears at the level of the couplings, it is expected to be largely 
independent of the particle species observed in the final state, with 
small corrections only due to e.g. the decay properties of heavy hadrons. 
In most fragmentation models, both for quark and gluon jet fragmentation,
the neutralisation of colour fields  occurs via the creation of 
quark-antiquark pairs. It has however been suggested~\cite{pw} that in 
the fragmentation of a gluon jet, colour neutralisation could also occur 
via the creation of gluon pairs.
In this case, an enhanced production of isoscalar mesons 
such as the $\eta$ would be expected in the fragmentation of gluon jets and, 
if they exist, also of glueballs. 

OPAL has measured $\pi^0$, $\eta$, $\rK^0$ and charged particle  
multiplicities in quark and gluon jets\cite{opal1} as a function of the 
hardness scale of the jet\cite{do} defined as 
$Q_{\jet}~\equiv~E_{\jet}\sin{\frac{\theta}{2}}$. 
From hadronic $\rZ^0$ decays with a 3-jet topology two samples of jets are 
selected consisting of the second and the third highest energy jet from 
each event, respectively. In the region $8$~GeV~$<~Q_{\jet}~<~26$~GeV, these 
samples contain different fractions of quark and gluon jets and the 
mean $\pi^0$, $\eta$, $\rK^0$ and charged particle multiplicities measured 
for the jets in these samples can be used to unfold the  
multiplicities for pure quark and gluon jets.

\begin{figure}[t] 
\begin{center}
\epsfxsize=13.85pc 
\epsfbox{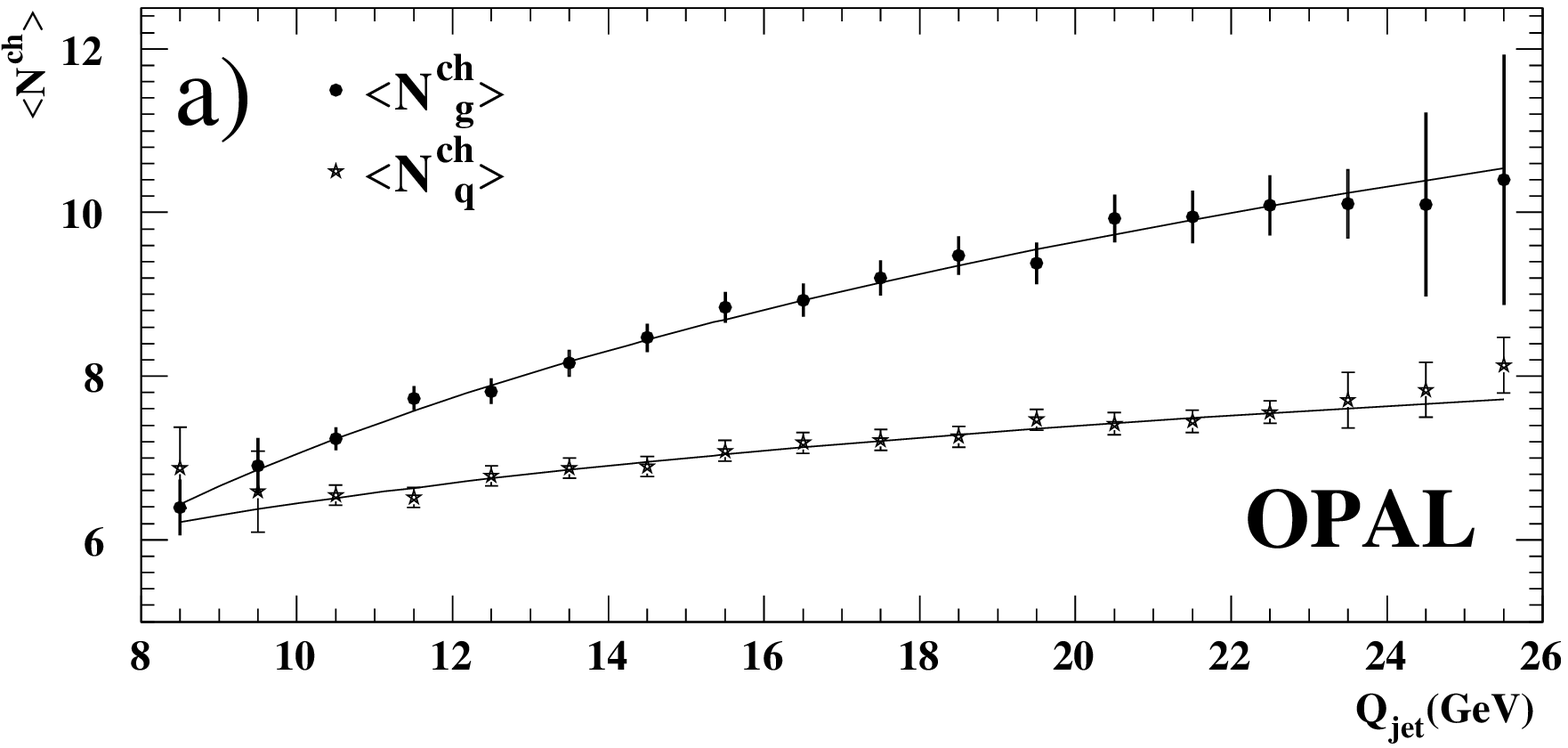} 
\epsfxsize=13.85pc 
\epsfbox{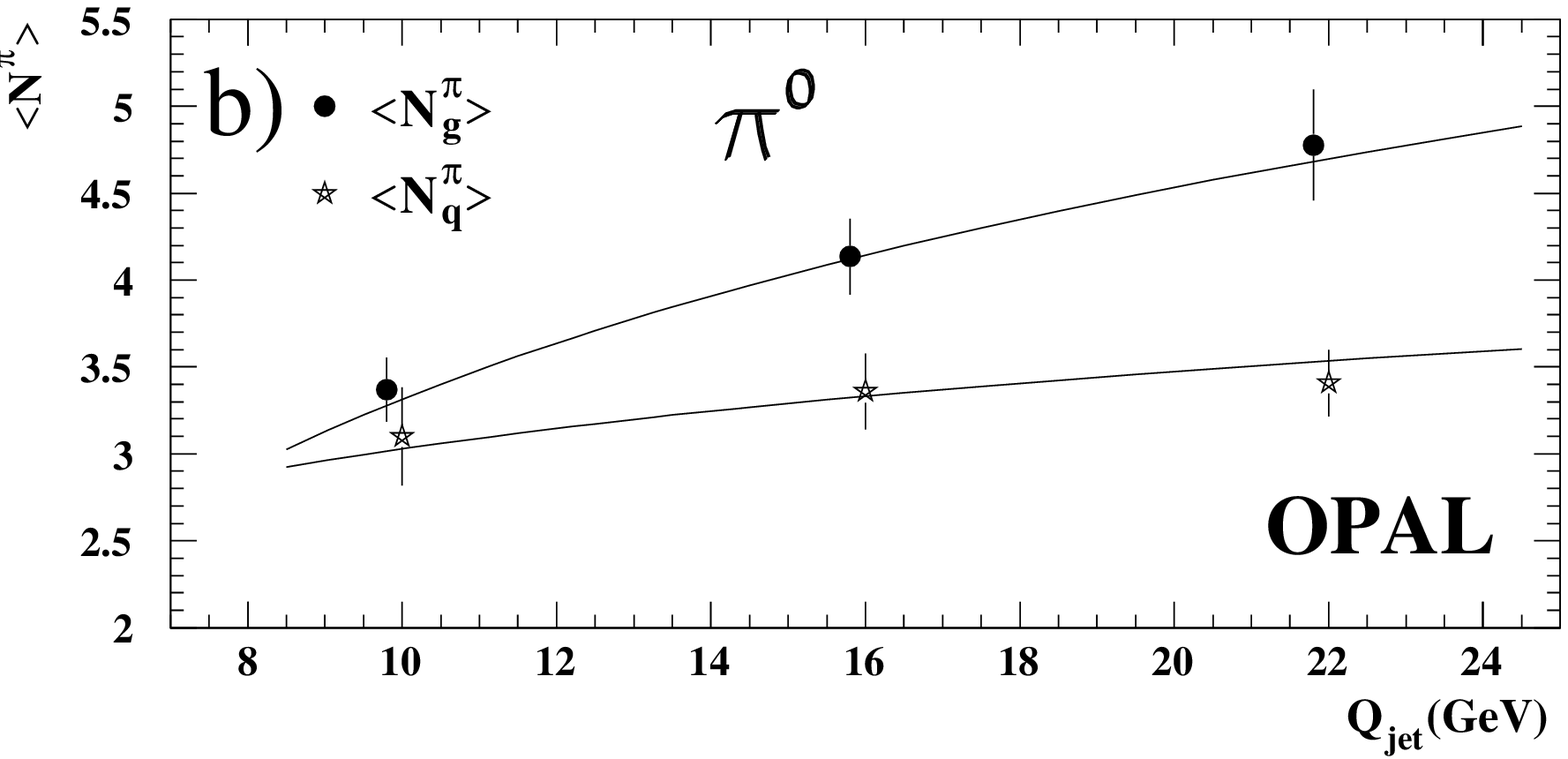} 
\epsfxsize=13.85pc 
\epsfbox{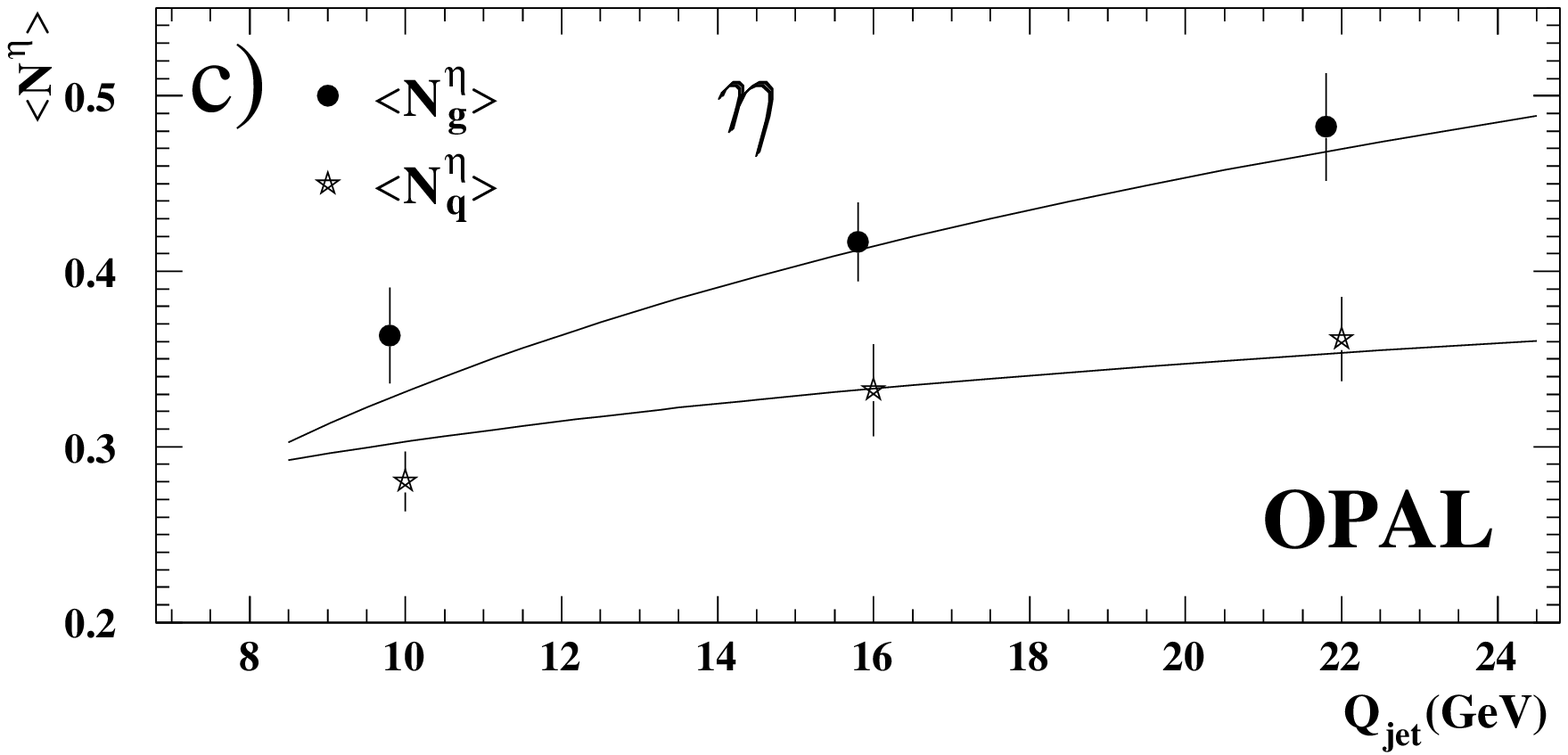} 
\epsfxsize=13.85pc 
\epsfbox{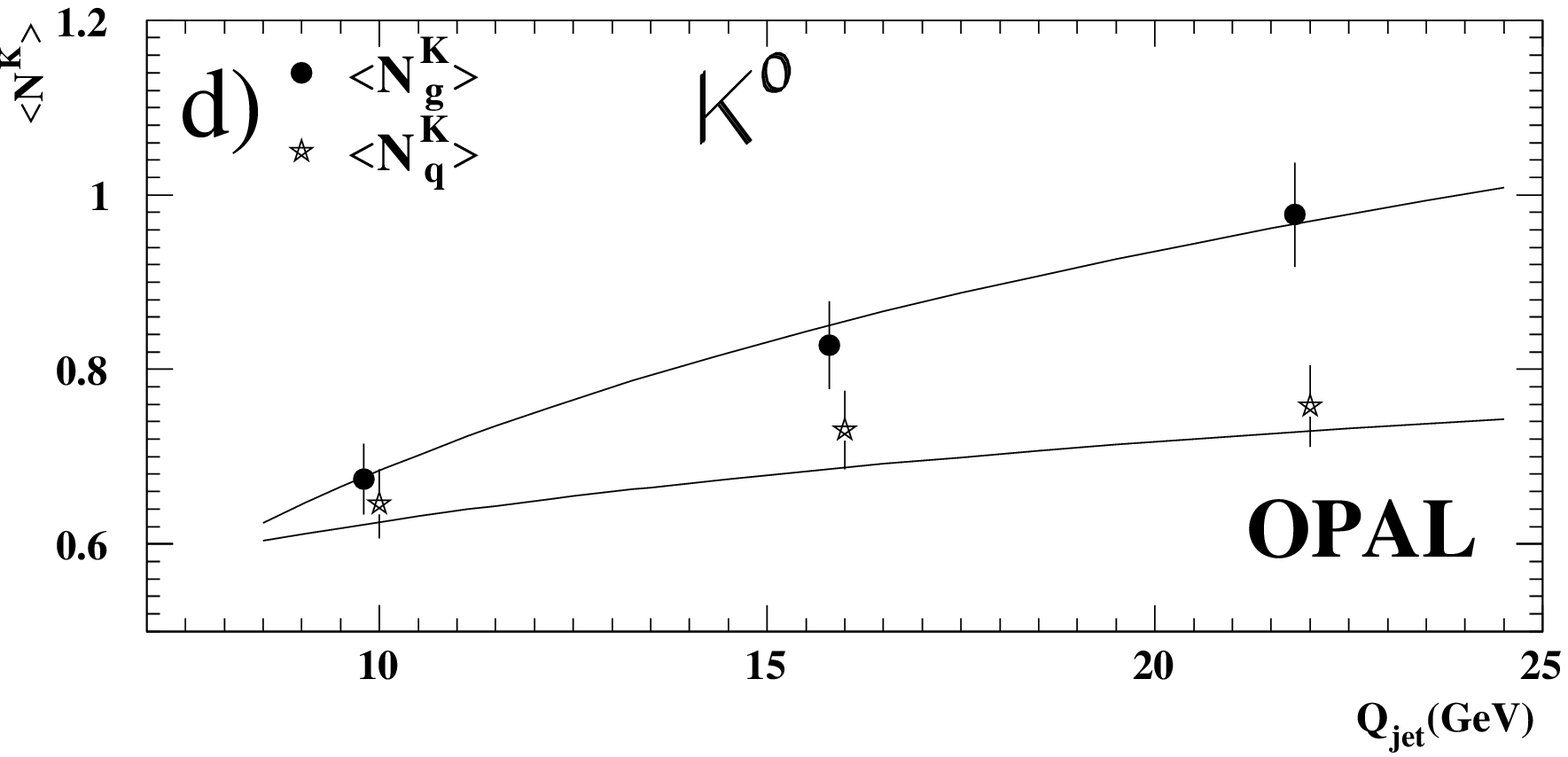}
\end{center} 
\caption{Mean charged particle, $\pi^0$, $\eta$ and $\rK^0$ multiplicity in quark and gluon jets, as a function of $Q_{\jet}$. The curves are the result of a fit to the charged particle multiplicity, scaled by a normalisation factor in the figures b,c and d.  \label{fig1}}
\end{figure}

In Fig.~\ref{fig1} we show the mean charged particle, $\pi^0$, 
$\eta$ and $\rK^0$ multiplicities in quark and gluon jets as 
a function of $Q_{\jet}$. Also shown is the result of a fit to the 
charged particle multiplicities, which has been scaled by an 
appropriate normalisation factor in figures b,c and d. 

\begin{figure}[t] 
\begin{center}
\epsfxsize=6.8pc 
\epsfbox{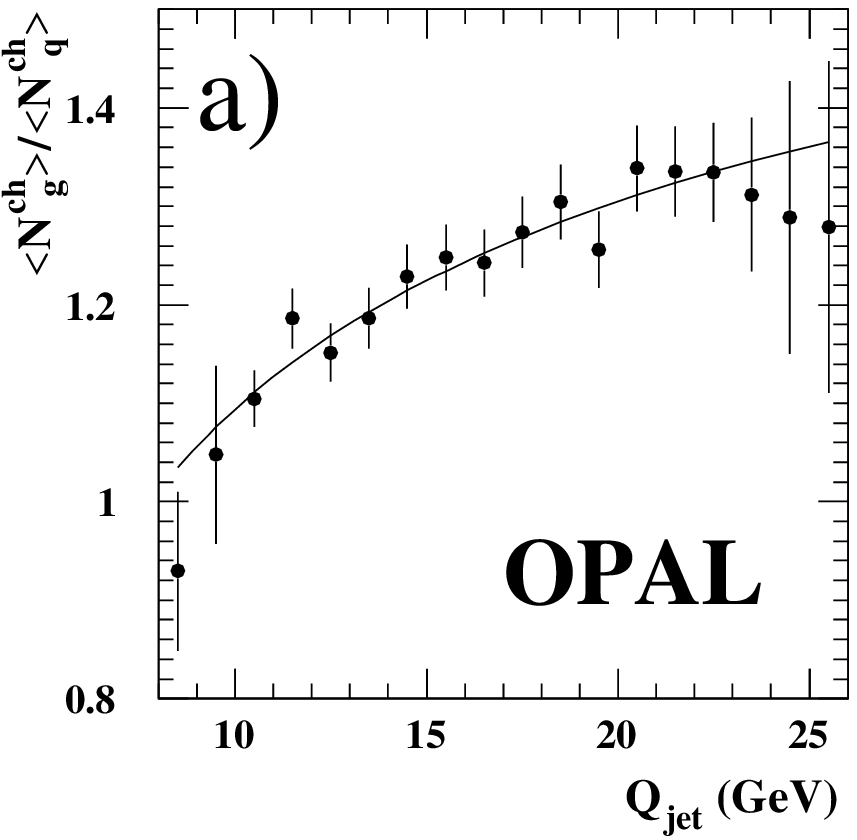} 
\epsfxsize=6.8pc 
\epsfbox{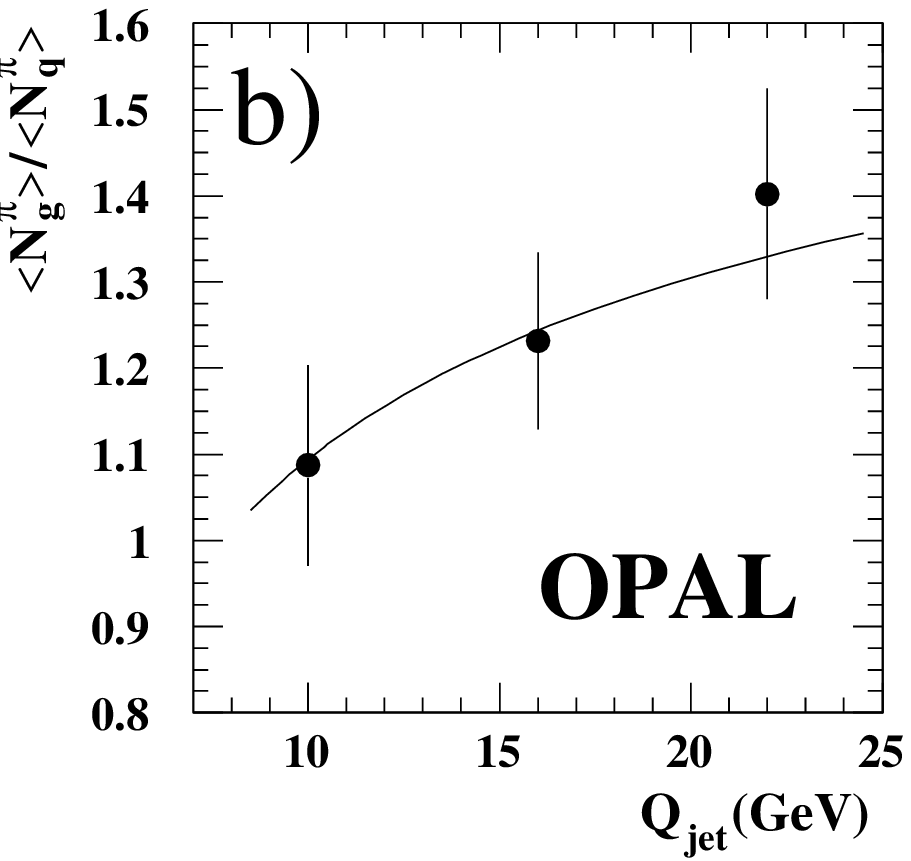} 
\epsfxsize=6.8pc 
\epsfbox{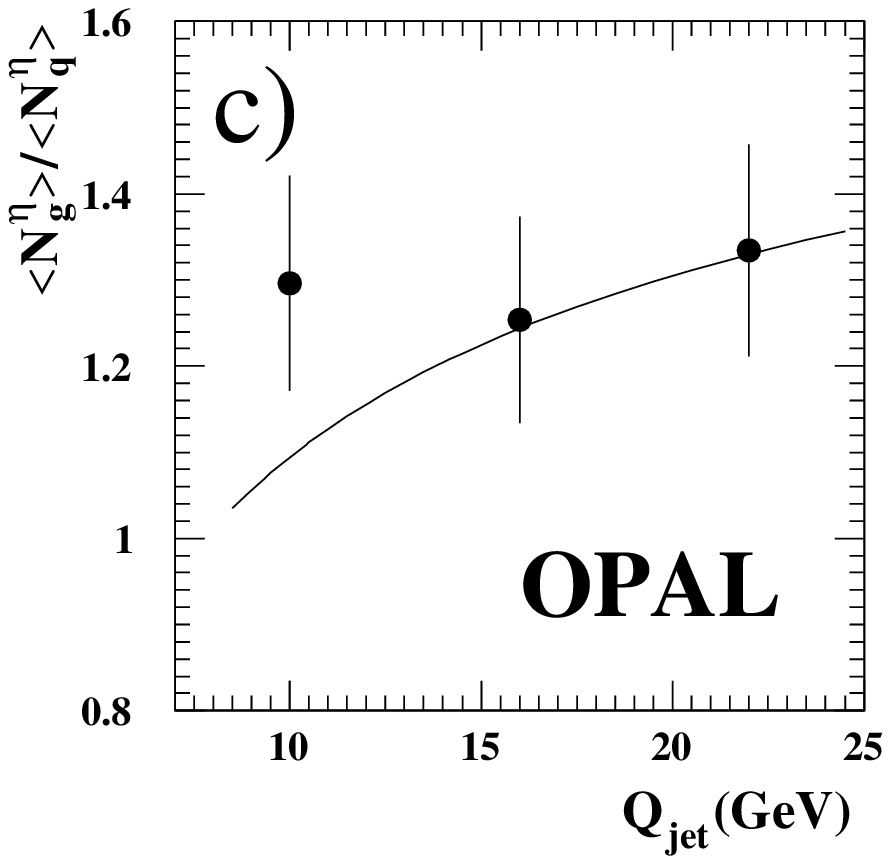} 
\epsfxsize=6.8pc 
\epsfbox{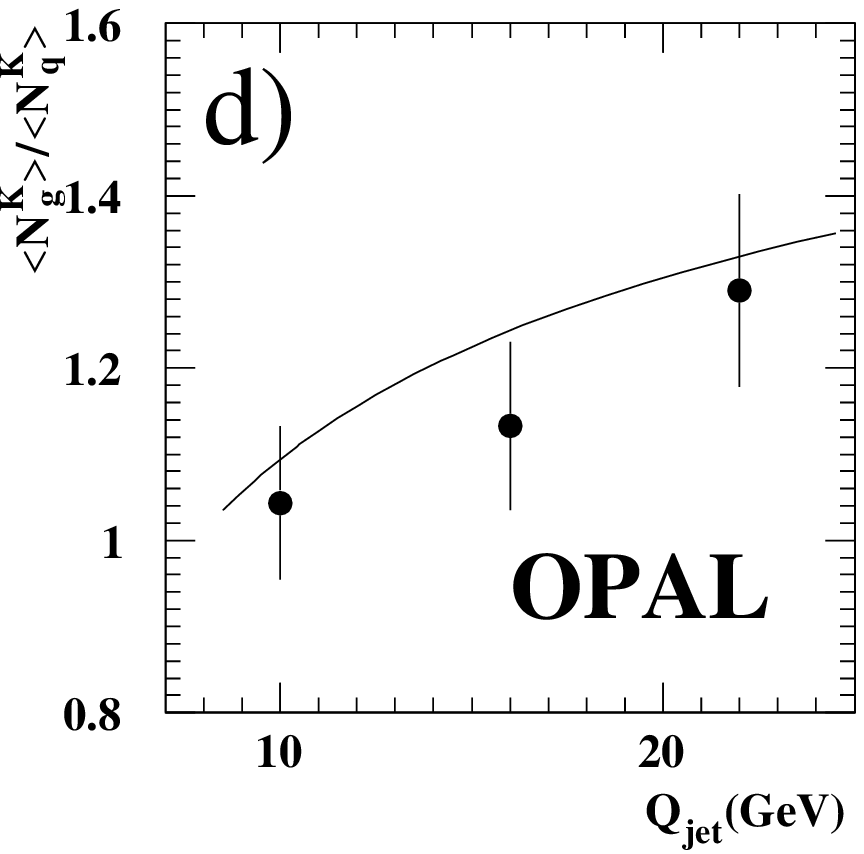}
\end{center} 
\caption{Ratio of the mean charged particle, $\pi^0$, $\eta$ and $\rK^0$ multiplicity in gluon and in quark jets as a function of $Q_{\jet}$. The curves are the result of a fit to the charged particle multiplicities.  \label{fig2}}
\end{figure}

Fig.~\ref{fig2} shows the ratio of the mean multiplicity in gluon and in 
quark jets for charged particles, $\pi^0$, $\eta$ and $\rK^0$. All 
cases can be described by the fit result obtained for charged particles. 
The multiplicity enhancement in gluon jets is thus found to be 
independent of the studied particle species.
These results do not confirm an enhanced $\eta$ production in three-jet 
events reported by the L3 collaboration.\cite{l3}

\section{\boldmath Leading particle production in light flavour jets}

OPAL has measured leading  $\pi^\pm$, $\rK^\pm$, $\rK^0_\rS$, 
$\rp(\bar{\rp})$ and $\Lambda(\bar{\Lambda})$ rates in up, down and strange 
flavour jets.\cite{opal2} As suggested in \cite{lpe}, these highest energy 
particles in a jet often carry the flavour of the quark from which the jet 
originated.
So far few experimental results\cite{sld} have confirmed  
this leading particle effect in the fragmentation of light flavour jets.
The method used by OPAL to determine leading particle rates in light flavour 
jets  was proposed in \cite{lm} and is based on counting all events in 
which a leading hadron is tagged in one of the event hemispheres and all  
events in which both 
hemispheres have a leading hadron tag. Using this information, our 
knowledge of the flavour composition of hadronic $\rZ^0$ decays and 
imposing some constraints based on isospin symmetry and flavour independence,  
the production rates for the different hadron species in 
up, down and strange flavour jets can be unfolded.

\begin{figure}[t] 
\begin{center}
\epsfxsize=13.85pc 
\epsfbox{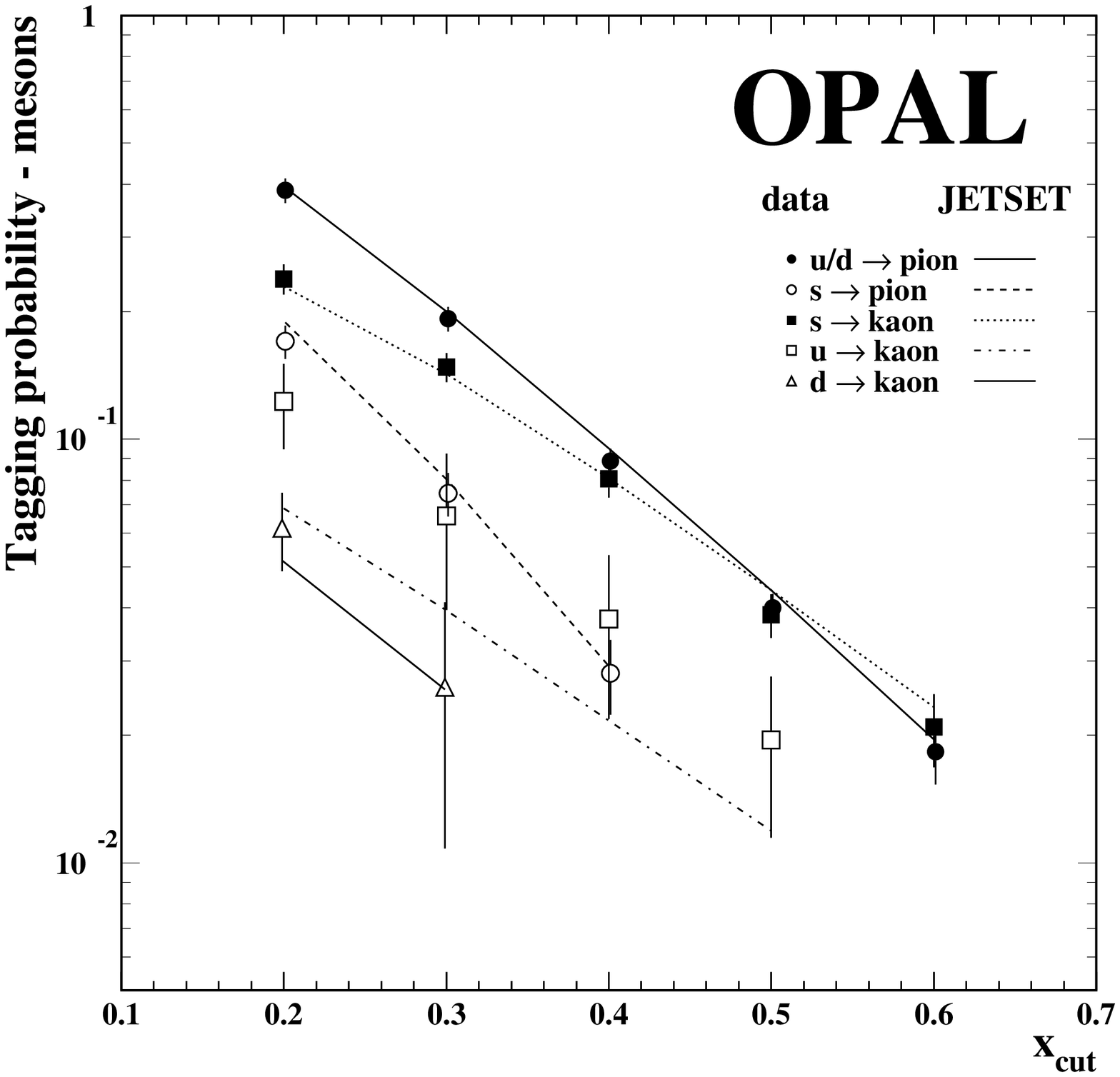} 
\epsfxsize=13.85pc 
\epsfbox{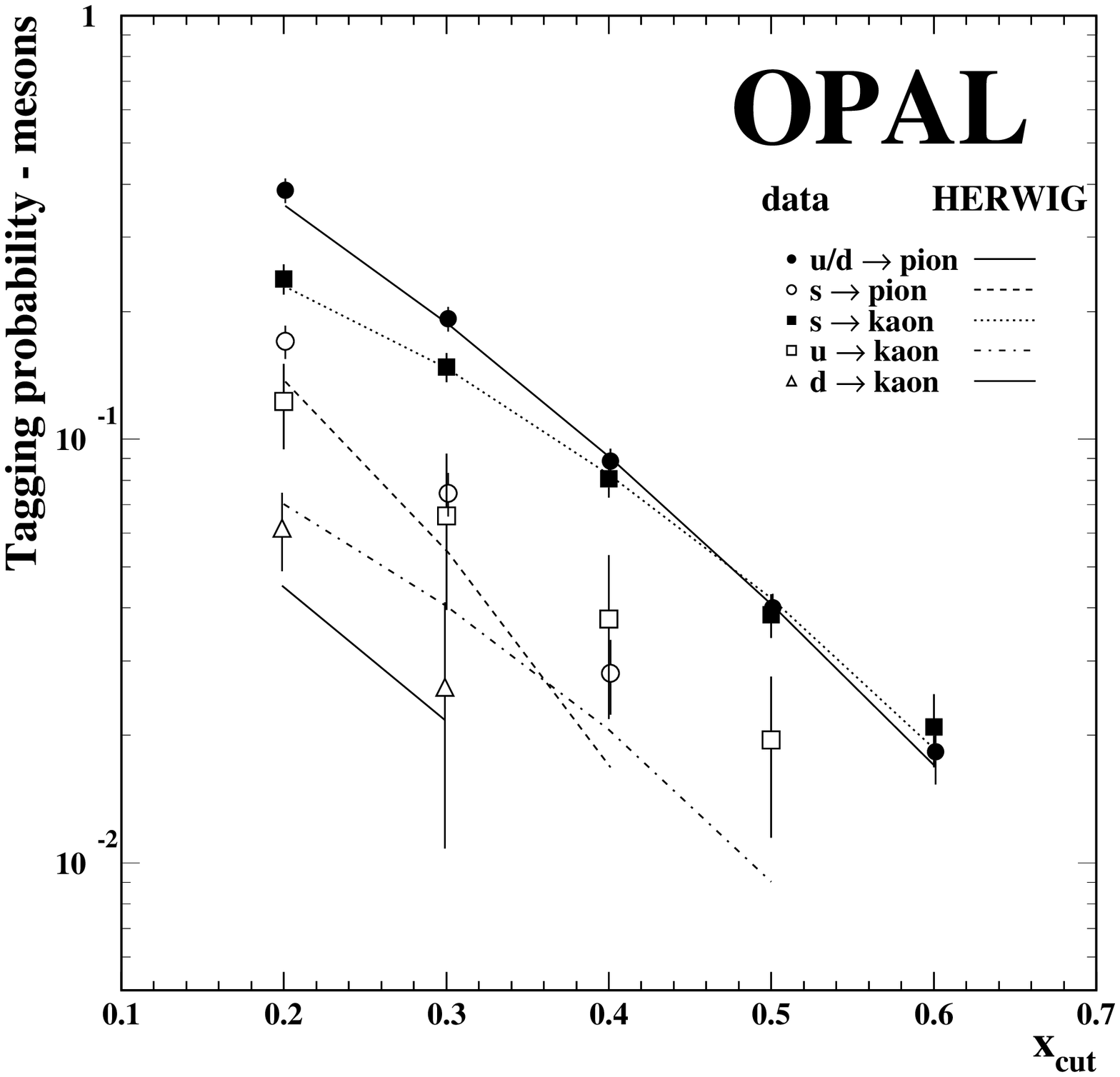} 
\end{center}
\caption{Tagging probabilities for pions and kaons in up, down and strange flavour jets, compared to JETSET (left) and HERWIG (right). \label{fig3}}
\end{figure}

\begin{figure}[b] 
\begin{center}
\epsfxsize=13.85pc 
\epsfbox{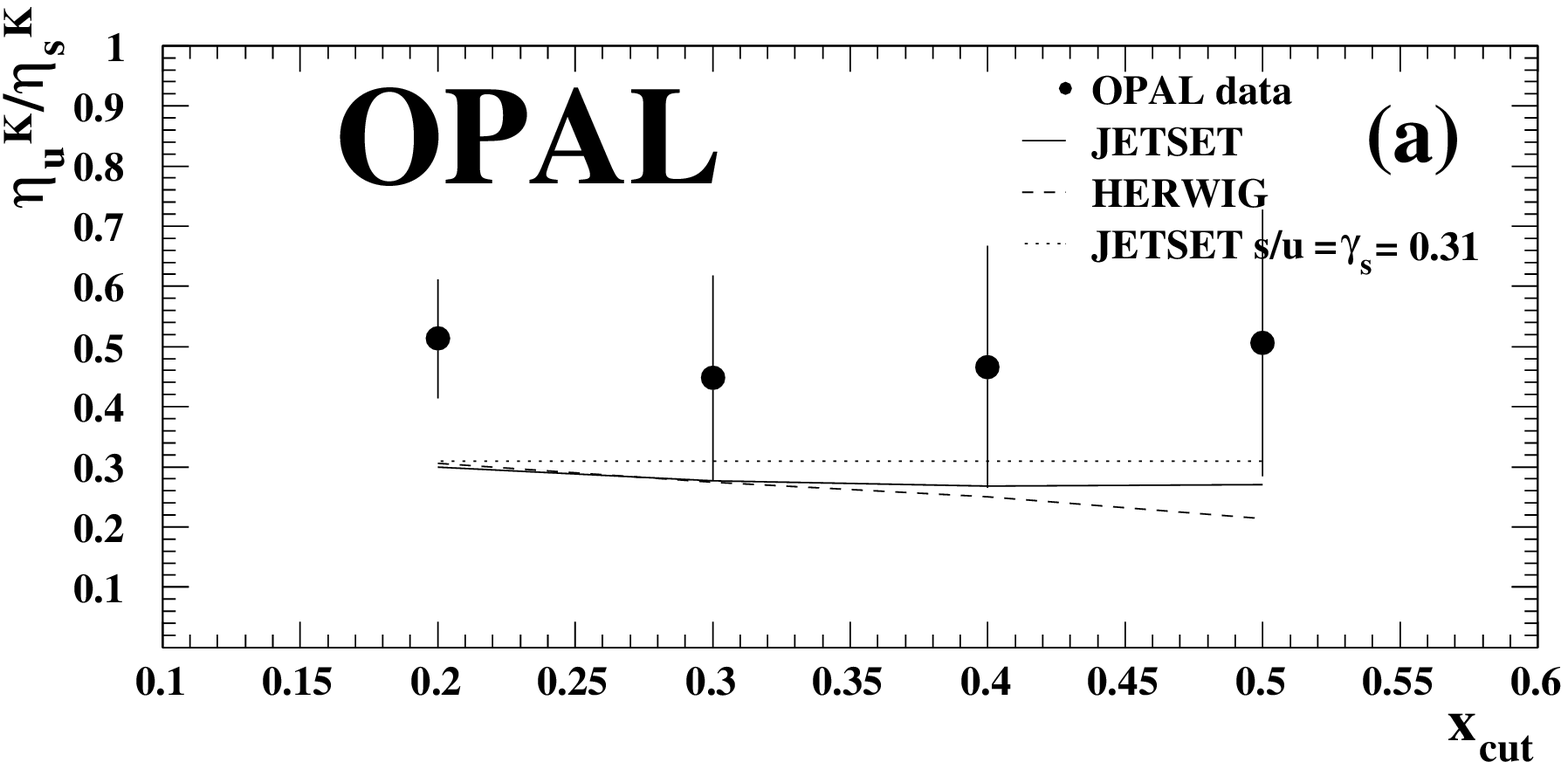} 
\epsfxsize=13.85pc 
\epsfbox{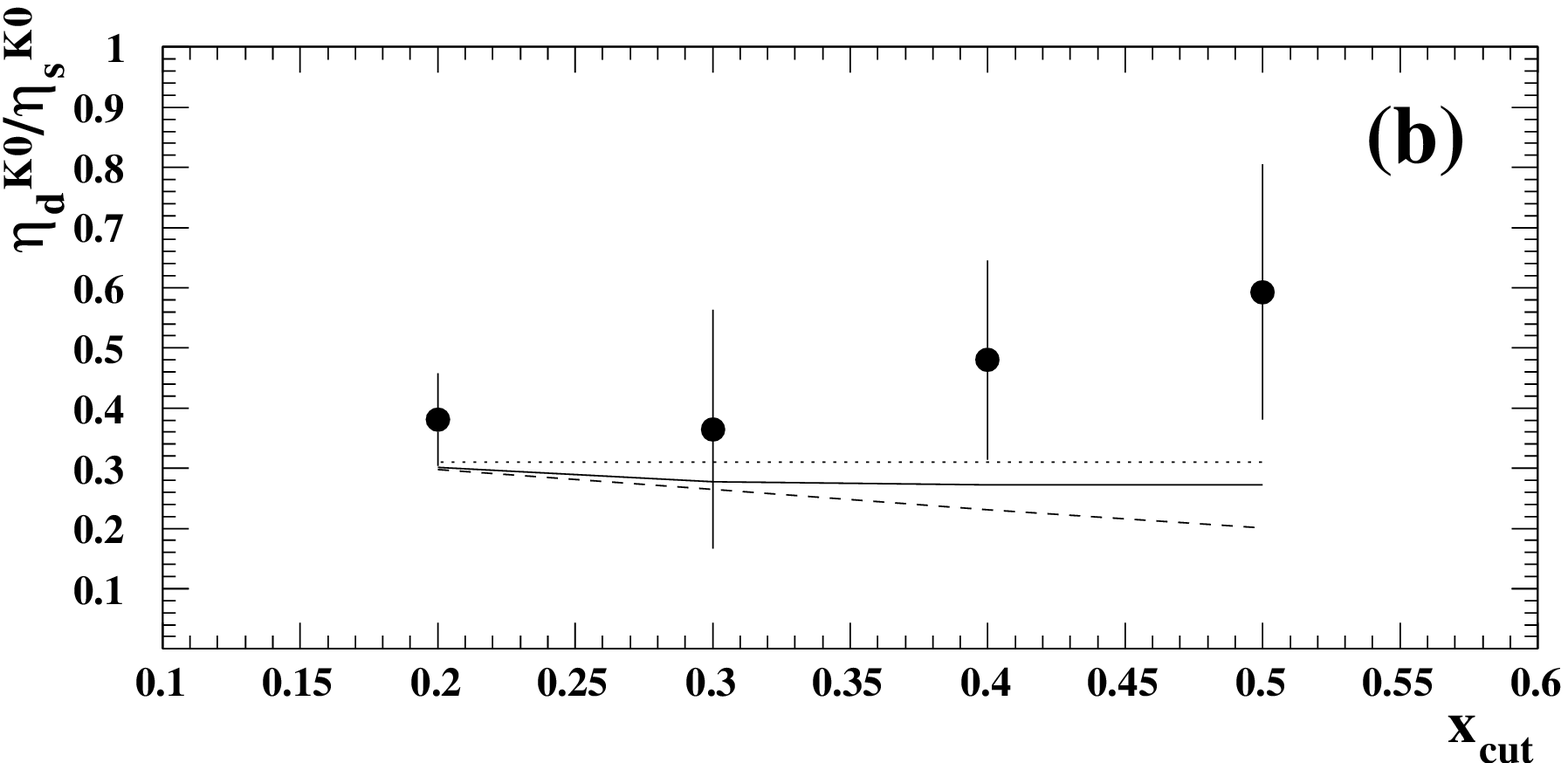} 
\end{center}
\caption{Ratio of tagging probabilities of $\rK^\pm$ in up and strange flavour jets (a) and of $\rK^0$ in down and strange flavour jets (b), compared to HERWIG and JETSET. The dash-dotted curves indicate the strangeness suppression factor ($\gamma_\rs=0.31$) used in JETSET. \label{fig4}}
\end{figure}

In Fig.~\ref{fig3}, the leading $\pi^\pm$ and $\rK^\pm$ rates in up, 
down and strange flavour jets are shown as a function of $x_{\cut}$, the lower 
threshold applied on the scaled momentum of the particle. 
The rates are compared to the corresponding JETSET\cite{jetset} and 
HERWIG\cite{herwig} predictions.
In accordance with the expected leading particle pattern, the highest rates 
are those of pions in up and down flavour jets and kaons in strange flavour 
jets.
The kaon rate in up flavour jets is 
lower than would na\"{\i}vely be expected from the leading particle effect. 
This is due to the suppressed production of strange quark-antiquark pairs 
from the hadronic sea. 
The strangeness suppression factor, associated with this effect, can be 
determined directly from the ratio of leading $\rK^\pm$ production in up and 
strange flavour jets and the ratio of leading $\rK^0$ production in down and 
strange flavour jets, shown in Fig.~\ref{fig4}a and \ref{fig4}b. 
The strangeness suppression factor obtained is 
\[ \gamma_\rs~=~0.422~\pm~0.049~({\rm stat.})~\pm~0.059~({\rm syst.}),\]
which is about one standard deviation high in comparison with other 
measurements,\cite{ssupp} in most of which $\gamma_\rs$ was determined 
rather indirectly, however.

\begin{figure}[t] 
\begin{center}
\epsfxsize=13.85pc 
\epsfbox{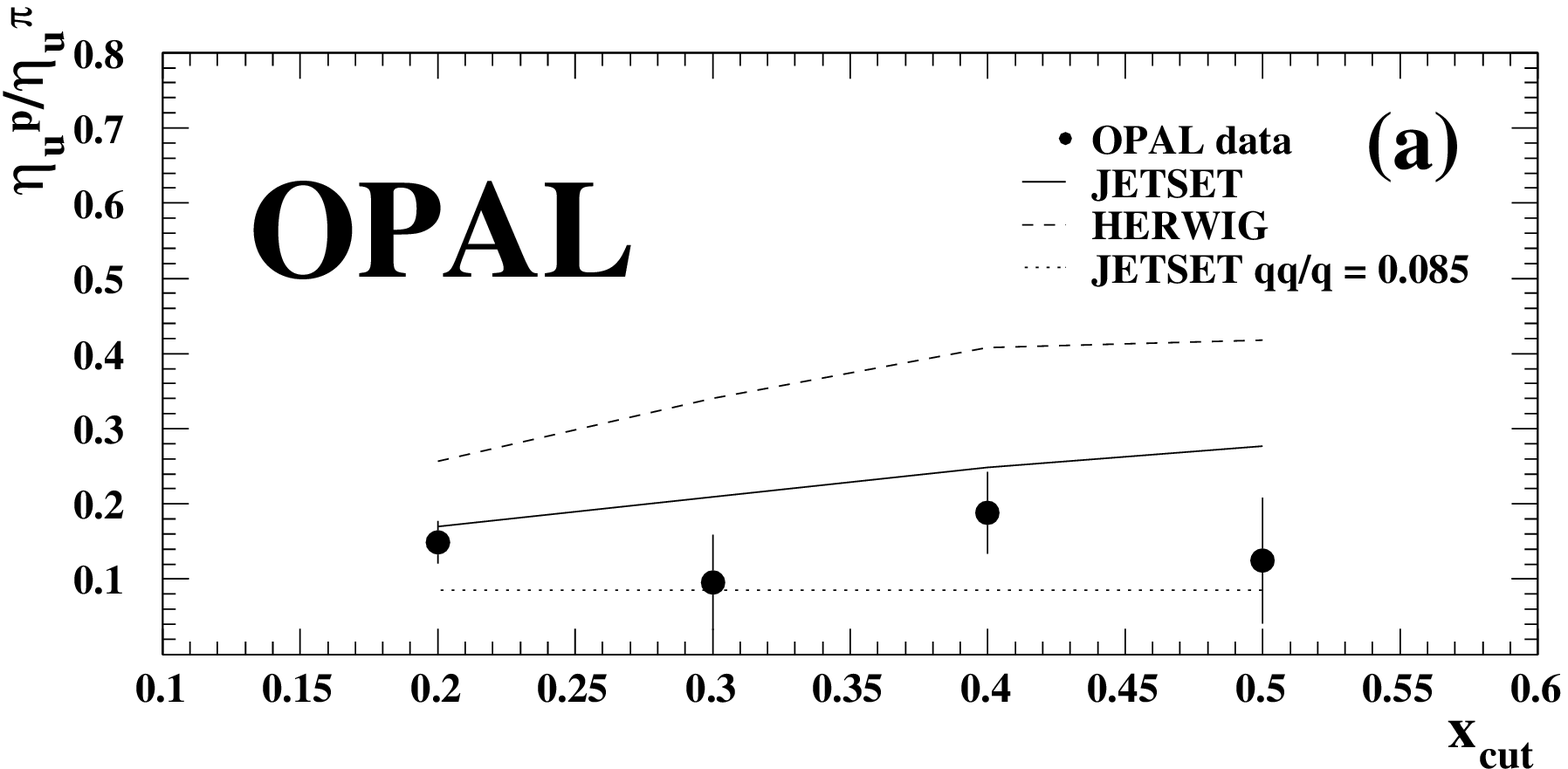} 
\epsfxsize=13.85pc 
\epsfbox{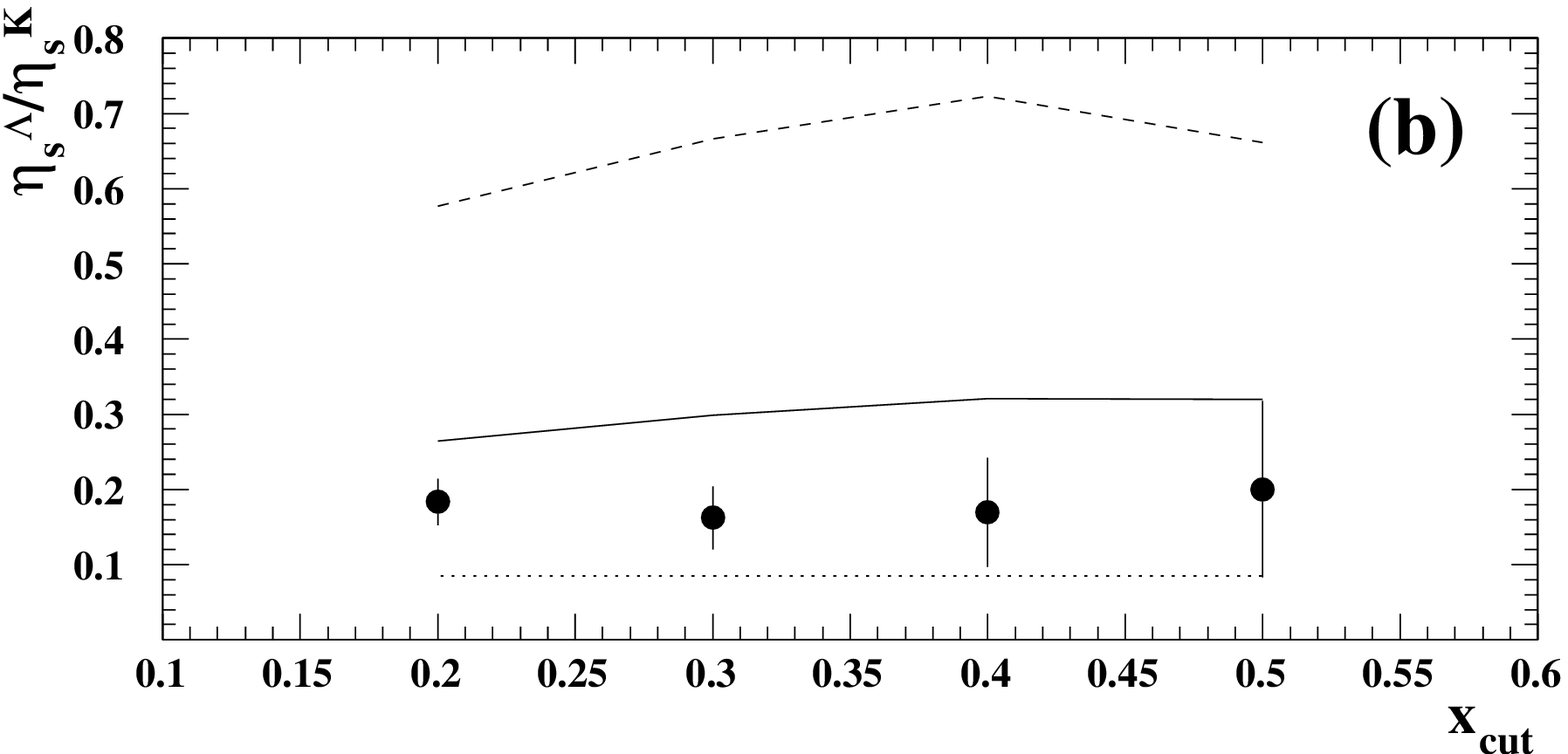} 
\end{center}
\caption{Ratio of tagging probabilities of protons and pions in up flavour jets (a) and of lambdas and kaons in down flavour jets (b), compared to HERWIG and JETSET. The dash-dotted curves indicate the diquark suppression factor 
($\rq\rq/\rq=0.085$) used in JETSET. \label{fig5}}
\end{figure}

We have also studied baryon production in the final state.
In the Monte Carlo models to which we compare our data, baryon production 
is governed by the creation of diquark pairs when breaking up strings
(in JETSET) or when decaying colourless clusters (in HERWIG).  
In Fig.~\ref{fig5}, the ratio of proton over pion production in up flavour 
jets and of lambda over kaon production in strange flavour jets are 
compared to JETSET and HERWIG. The HERWIG Monte Carlo significantly 
overestimates the fraction of baryons produced. 
 
\section{\boldmath Charged particle multiplicity in up, down and strange flavoured $\rZ^0$ decays}

OPAL has studied mean charged particle multiplicities in up, down and 
strange flavoured $\rZ^0$ decays.\cite{opal3} Because of the flavour 
independence of the strong interaction, these multiplicities are expected 
to be identical, with small corrections only due to the decay properties 
of heavy hadrons.

For the purpose of the measurement, the charged particle multiplicity has been
determined for three event samples, defined by the observation of 
a leading $\rK^\pm$ with fractional energy $x_E>0.5$, of a leading $\rK^0_\rS$ 
with $x_E>0.4$, or of a leading charged hadron with $x_E>0.7$, respectively. 
These samples 
differ in the relative fractions of up, down and strange flavoured events 
they contain and have been used to unfold the mean charged particle 
multiplicity for purely up, down or strange flavoured $\rZ^0$ decays. 
These are found to be 
\[\langle n_\ru\rangle~=~17.77~\pm~0.51~({\rm stat.})~^{+0.86}_{-1.20}~({\rm syst.}),\]
\[\langle n_\rd\rangle~=~21.44~\pm~0.63~({\rm stat.})~^{+1.46}_{-1.17}~({\rm syst.}),\]
\[\langle n_\rs\rangle~=~20.02~\pm~0.13~({\rm stat.})~^{+0.39}_{-0.37}~({\rm syst.}),\]
where we point out that in particular the results for up and down 
flavoured events are strongly anti-correlated. Within the uncertainties, the 
measured mean multiplicities are consistent with being identical, as 
expected from the flavour independence of the strong interaction.

\section{\boldmath Conclusions}

Mean $\pi^0$, $\eta$, $\rK^0$ and charged particle multiplicities 
have been determined for quark and gluon jets. The multiplicity 
enhancement in gluon jets is found to be independent of the 
studied particle species. 

We have measured leading hadron production rates for various hadron 
species in up, down and strange flavour jets. The results confirm the 
leading particle effect in the fragmentation of light flavour jets and 
have been used for a direct determination of the strangeness suppression 
factor, yielding $\gamma_\rs~=~0.422~\pm~0.049~({\rm stat.})~\pm~0.059~({\rm syst.})$.
In addition the results constitute a detailed test of fragmentation models. 
In particular the HERWIG Monte Carlo model is found to strongly overestimate 
baryon production in the final state.

The mean charged particle multiplicity has been determined for up, down and 
strange flavoured $\rZ^0$ decays. The results are identical within the 
uncertainties of the measurement, as expected from the flavour 
independence of the strong interaction.

\def\EPJ{{\em Eur. Phys. J.} C}
\def\CPC{{\em Comp. Phys. Comm.}}

\end{document}
